\documentclass[preprintnumbers,amsmath,11pt,amssymb,floatfix,superscriptaddress,noshowpacs,nofootinbib]{revtex4}
\usepackage{graphicx}
\usepackage{epsfig}
\usepackage{bm}
\usepackage{amsfonts}

\input epsf

\begin{document}
\title{\Large A dark energy model with higher order derivatives of $H$ in the $f(R,T)$ modified gravity model}

\author{Antonio Pasqua}
\email{toto.pasqua@gmail.com} \affiliation{Department of Physics,
University of Trieste, Via Valerio, 2 34127 Trieste, Italy.}

\author{Surajit Chattopadhyay}
\email{surajit_2008@yahoo.co.in, surajcha@iucaa.ernet.in}
\affiliation{ Pailan College of Management and Technology, Bengal
Pailan Park, Kolkata-700 104, India.}

\author{Ratbay Myrzakulov}
\email{rmyrzakulov@gmail.com}
\affiliation{Eurasian International Center for Theoretical Physics, Eurasian National University, Astana, 010008, Kazakhstan.}

\date{\today}

\begin{abstract}
\textbf{Abstract}:  In this paper, we consider a model of Dark
Energy (DE)  which contains three terms (one proportional to the squared Hubble parameter, one to the first derivative and one to the second  derivative with respect to the cosmic time of the Hubble parameter)  in the light of the $f\left(R,T\right) = \mu R + \nu T$ modified gravity model, with $\mu$ and $\nu$ being two
constant parameters. $R$ and $T$ represent the curvature and torsion scalars, respectively.
We found that the Hubble parameter exhibits a decaying behavior until redshifts  $z\approx-0.5$ (when it starts to increase) and the time derivative of the Hubble parameter goes from negative to positive values for different redshifts. The equation of state (EoS)
parameter of DE and the effective EoS parameter exhibit a transition from $\omega<-1$ to $\omega>-1$  (showing a quintom-like behavior). We also found that the model considered can attain the late time accelerated phase of the universe. Using the statefinder parameters $r$ and $s$, we derived that the studied model  can
attain the $\Lambda$CDM phase of the universe and can
interpolate between dust and $\Lambda$CDM phase of the universe.
Finally, studying the squared speed of sound $v_s^2$, we found that the considered model is classically stable in the earlier stage of the
universe, but classically unstable in the current stage.
\end{abstract}

\maketitle

\section{Introduction}
The late-time accelerated expansion of the universe (which is well-established from different cosmological observations) \cite{spe,perlm} is a major challenge for cosmologists. The
universe underwent two phases of accelerated expansion: the
inflationary stage in the very early universe and a late-time
acceleration in which our universe entered only recently. Models trying to explain this late-time acceleration
are dubbed as Dark Energy (DE) models. An important step toward the comprehension of
the nature of DE is to understand whether it is produced by a
cosmological constant $\Lambda$ or it is originated from other
sources dynamically changing with time \cite{tsu}. For good reviews on DE see
\cite{cople1,pad,bamba1}.

 In a recent paper,
Nojiri $\&$ Odintsov \cite{nogiri} described the reasons why
modified gravity approach is extremely attractive in the
applications for late accelerating universe and DE. Another good
review on modified gravity was made by Clifton et al. \cite{cli}.
Many different theories of modified gravity have been recently proposed:
some of them are $f\left(R\right)$ (with $R$ being the Ricci
scalar curvature) \cite{nojiri2,nojiri1}, $f\left( T\right)$ (with
$T$ being the torsion scalar)
\cite{cai,fer,bamba2,bamba3}, Ho\v{r}ava-Lifshitz \cite{kir,nis} and
Gauss-Bonnet \cite{myr22,ban,nojiri3,li} theories.

In this paper, we
concentrate on $f\left(R, T\right)$ gravity, with $f$ being in
this case a function of both $R$ and $T$, manifesting a coupling
between matter and geometry. Before going into the details of
$f\left(R, T\right)$ gravity, we describe some important features
of the $f\left(R\right)$ gravity. The recent motivation for
studying $f\left(R\right)$ gravity came from the necessity to
explain the apparent late-time accelerating expansion of the
Universe. Detailed reviews on $f\left(R\right)$ gravity can be
found in \cite{sot1,defelice,sot2,cap}. Thermodynamic aspects of
$f\left(R\right)$ gravity have been investigated in the works of
\cite{bamba4,akbar}. A generalization of $f\left(R\right)$ modified
theory of gravity including in the theory an explicit coupling of an arbitrary function of $R$ with
the matter Lagrangian density $L_m$ leads to the motion of massive particles is non-geodesic, and an
extra force, orthogonal to the four-velocity, arises \cite{pop}. Harko et al. \cite{har} recently suggested
an extension of standard General Relativity, where the
gravitational Lagrangian is given by an arbitrary function of $R$
and $T$ and called this model as $f\left(R, T\right)$. The
$f\left(R, T\right)$ model depends on a source term,
representing the variation of the matter stress-energy tensor with
respect to the metric. A general expression for this source term
can be obtained as a function of the matter Lagrangian $L_m$. In a recent paper, Myrzakulov \cite{myr1} proposed $f\left( R, T\right)$ gravity model and studied its  main properties of FRW cosmology. Moreover,  Myrzakulov \cite{myr2}
recently derived exact solutions for a specific
$f\left(R,T\right)$ model which is a linear combination of $R$ and
$T$, i.e. $f\left(R, T\right) = \mu R+\nu T$, where $\mu$ and
$\nu$ are two free constant parameters. Moreover, it was demonstrated that, for
some specific values of $\mu$ and $\nu$, the expansion of universe
results to be accelerated without the necessity to introduce extra dark
components. Recently, Chattopadhyay \cite{chatto} studied the properties of interacting Ricci DE considering the model $f\left(R, T\right) = \mu R+\nu T$. Pasqua et al. \cite{miofrt} recently considered the Modified Holographic Ricci Dark Energy (MHRDE) model in the context of the specific $f\left(R,T\right)$ model we are considering in this work.
Moreover, Alvarenga et al. \cite{alva} studied the evolution of scalar cosmological perturbations in the metric formalism in the framework of $f\left(R,T\right)$ modified theory of gravity.\\
In this work, we consider a DE model proposed in the recent paper of Chen $\&$ Jing \cite{chens}. The DE model considered contains three different term, one proportional one proportional to the squared Hubble parameter, one to the first derivative with respect to the cosmic time of the Hubble parameter and one proportional to the second  derivative with respect to the cosmic time of the Hubble parameter:
\begin{eqnarray}
    \rho_{DE} =  \varepsilon \frac{\ddot{H} }{H} + \lambda \dot{H}  +\theta H^2  , \label{enerden}
\end{eqnarray}
where $\varepsilon$, $\lambda$ and $\theta$ are three positive constant parameters. The first term is divided by the Hubble parameter $H$ in order all the three terms have the same dimensions. The energy density given in Eq. (\ref{enerden}) can be considered as an extension and generalization of other two DE models widely studied in recent time, i.e. the Ricci DE (RDE) model and the the DE energy density with Granda - Oliveros cut-off. In fact, in the limiting case corresponding to $\varepsilon =0$, we obtain the energy density of DE with Granda-Oliveros cut-off, and in the limiting case corresponding to $\varepsilon=0$, $\lambda = 1$ and $\theta =2$, we recover the RDE model for flat universe (i.e. with curvature parameter $k$ equal to zero).\\
In this work we are considering DE interacting with pressureless DM which has energy density $\rho_m$. Various form of interacting DE models have been constructed in order to fulfil the observational requirements. Many different works are presently available where the interacting DE have been discussed in details. Some examples of interacting DE are presented in \cite{JAMM,wu2,kim2,setmr,wa,karami}.\\
This work aims to reconstruct the DE model considered  under $f\left(R, T\right)$ gravity and it  is organized as follow. In Section 2, we describe the main features of the $f\left(R,T\right) = \mu R + \nu T$ model. In Section 3, we consider the energy density of DE given in Eq. (\ref{enerden}) in the context of $f\left(R,T\right) $ gravity considering the particular model considered. In Section 4, we study the statefinder parameters $r$ and $s$ for the energy density model we are considering in this work. In Section 5, we write a detailed discussion about the results found in this work.  Finally, in Section  6, we write the Conclusions of this work.

\section{The $f\left(R,T\right) = \mu R + \nu T$ model}
The metric of a spatially flat, homogeneous and isotropic universe in Friedmann-Lemaitre-Robertson-Walker (FLRW) model is given by:
\begin{eqnarray}
    ds^2=dt^2-a^2\left(t\right)\left[dr^2 +r^2 \left(d\theta ^2 + \sin^2 \theta d\varphi ^2\right) \right], \label{2}
\end{eqnarray}
where $a\left(t\right)$ represents a dimensionless scale factor (which gives information about the expansion of the Universe), $t$ indicates the cosmic time, $r$ represents the radial component and $\theta$ and $\varphi$ are the two angular coordinates. Since we are dealing with flat metric, the curvature parameter $k$ is taken equal to zero $\left( k=0\right)$. \\
We also know that the tetrad orthonormal components $e_i\left( x^{\mu}  \right)$ are related to the metric through the following relation:
\begin{eqnarray}
g_{\mu \nu} = \eta _{ij}e^i_{\mu} e^j_{\nu} \label{metric}
\end{eqnarray}
The Einstein field equations are given by:
\begin{eqnarray}
H^2 &=& \frac{1}{3}\rho,\label{3} \\
\dot{H} &=& -\frac{1}{2}\left(\rho + p  \right), \label{4}
\end{eqnarray}
where $\rho$ and $p$ indicate (choosing units of $8 \pi G = c = 1$) the total energy density and the total pressure, respectively . The conservation equation is given by:
\begin{eqnarray}
\dot{\rho}+3H\left(\rho + p\right)=0, \label{5}
\end{eqnarray}
where:
\begin{eqnarray}
\rho &=& \rho_{DE} + \rho_m,\label{6} \\
p &=& p_{DE}. \label{7}
\end{eqnarray}
We must emphasize here that we are considering pressureless DM ($p_m = 0$). Since the components
do not satisfy the conservation equation separately in presence of interaction, we reconstruct
the conservation equation by introducing an interaction term $Q$ which can be expressed in
any of the following forms \cite{shei}: $Q  \propto  H\rho_{DE}$, $Q \propto  H\rho_m$ and $Q \propto  H\left(\rho_m + \rho_{DE}\right)$. In this paper, we consider+ as interaction term the second of the three forms mentioned above. Accordingly, the conservation equation is reconstructed as:
\begin{eqnarray}
\dot{\rho}_{DE}+3H\left(\rho_{DE} + p_{DE}\right) &=& 3H\delta \rho_m, \label{8}  \\
\dot{\rho}_m+3H\rho_m &=& -3H\delta \rho_m, \label{9}
\end{eqnarray}
where $\delta$ indicates an interaction constant parameter which gives information about the strength of the interaction between DE and DM. The present day value of $\delta$ is still not known exactly and it is under debate.\\
One of the most interesting models of $f\left(R, T\right)$ gravity is the so-called $M_{37}$-model, which action $S$ is given by \cite{myr1}:
\begin{eqnarray}
S = \int f\left(  R,T \right)ed^4x + \int L_med^4x, \label{10}
\end{eqnarray}
where $e$ is defined as $e=det\left( e^i_{\mu} \right)=\sqrt{-g}$ (with $g$ being the determinant of the metric tensor $g_{\mu \nu}$), $L_m$ is the matter Lagrangian, $R$ is the curvature scalar and $T$ is the torsion scalar. \\
In this paper, we consider the following expressions for the curvature scalar $R$ and for the torsion scalar $T$, respectively:
\begin{eqnarray}
R &=& u + 6\left( \dot{H} + 2H^2  \right), \label{11}\\
T &=& v - 6H^2. \label{12}
\end{eqnarray}
We now consider the particular case corresponding to $u = u\left( a, \dot{a}  \right)$ and $v = v\left( a, \dot{a}  \right)$, where $\dot{a}$ is the derivative of the scale factor with respect to the cosmic time $t$. Moreover, the scale factor $a\left(t\right)$,
the torsion scalar $T$ and the curvature scalar $R$ are considered as independent dynamical variables.
Then, after some algebraic calculations, the action given in Eq. (\ref{10}) can be rewritten as:
\begin{eqnarray}
S_{37} = \int dt L_{37},
\end{eqnarray}
where the Lagrangian $L_{37}$ is given by:
\begin{eqnarray}
L_{37} &=& a^3 \left( f -T f_T -Rf_R +v f_T + uf_R  \right)\\
&&- 6\left(f_R + f_T   \right)a\dot{a}^2  - 6\left(f_{RR} \dot{R} + f_{RT}\dot{T}   \right)a^2\dot{a} -a^3L_m.
\end{eqnarray}
The quantities $f_R$, $f_T$, $f_{RR}$ and $f_{RT}$ are, respectively, the first derivative of $f$ respect to $R$, the first derivative of $f$ respect to $T$, the second derivative of $f$ respect to $R$ and the second derivative of $f$ respect to $R$ and $T$.\\
The equations of $f\left(R, T\right)$ gravity are usually more complicated with respect to the equations of Einstein's Theory of General Relativity even if the FLRW metric is considered. For this reason, as stated before, we consider the following simple particular model of $f\left(R, T\right)$ gravity:
\begin{eqnarray}
f\left(  R, T \right) = \mu R + \nu T, \label{13}
\end{eqnarray}
with $\mu$ and $\nu$ two constant parameters.\\
The equations system of this model of $f\left(R, T\right)$ gravity is given by:
\begin{eqnarray}
\mu D_1 + \nu E_1 + K \left( \mu R + \nu T   \right) &=& -2a^3 \rho, \label{14}\\
\mu A_1 + \nu B_1 + M \left( \mu R + \nu T   \right) &=& 6a^2 p, \label{15}\\
\dot{\rho} + 3H\left( \rho + p  \right) &=& 0,\label{16}
\end{eqnarray}
where:
\begin{eqnarray}
D_1 &=& -6a\dot{a}^2+ a^3u_{\dot{a}}\dot{a}-a^3\left( u-R  \right)  = 6a^2\ddot{a} +a^3\dot{a}u_{\dot{a}}  = a^3\left( 6\frac{\ddot{a}}{a} + \dot{a}u_{\dot{a}} \right),\label{17} \\
E_1 &=&  -6a\dot{a}^2+a^3\dot{a}v_{\dot{a}} -a^3\left( v-T  \right) = -12a\dot{a}^2+a^3\dot{a}v_{\dot{a}} =   a^3\left( -12\frac{\dot{a}^2}{a^2} + \dot{a}v_{\dot{a}}\right), \label{18}\\
K &=& - a^3, \label{19}\\
A_1 &=& 12\dot{a}^2 + 6a\ddot{a} + 3a^2\dot{a}u_{\dot{a}}+a^3u_{\dot{a}} -a^3u_{a},\label{20} \\
B_1 &=& -24\dot{a}^2 -12a\ddot{a} + 3a^2\dot{a}v_{\dot{a}}+a^3v_{\dot{a}} -a^3v_{a},\label{21}\\
M &=& -3a^2.\label{22}
\end{eqnarray}
We get from Eqs. (\ref{14}), (\ref{15}) and (\ref{16}):
\begin{eqnarray}
-6\left( \mu + \nu \right) \frac{\dot{a}^2}{a^2} + \mu \dot{a}u_{\dot{a}} +\nu \dot{a}v_{\dot{a}} -\mu u -\nu v &=& -2\rho, \label{AAA}\\
-2\left( \mu + \nu   \right)\left(\frac{\dot{a}^2}{a^2} + 2\frac{\ddot{a}}{a}   \right)      + \mu \dot{a}u_{\dot{a}} +\nu \dot{a}v_{\dot{a}} -\mu u -\nu v + \frac{\mu}{3}a\left(\dot{u}_{\dot{a}}-u_a\right) + \frac{\nu}{3}a\left(\dot{v}_{\dot{a}}-v_a\right) &=& 2p, \label{BBB}\\
\dot{\rho} + 3H\left( \rho + p  \right) &=& 0.  \label{CCC}
\end{eqnarray}
Then, Eqs. (\ref{AAA}), (\ref{BBB}) and (\ref{CCC}) can be rewritten as follow:
\begin{eqnarray}
3\left( \mu + \nu \right) \frac{\dot{a}^2}{a^2} -\frac{1}{2}\left( \mu \dot{a}u_{\dot{a}} +\nu \dot{a}v_{\dot{a}} -\mu u -\nu v\right) &=& \rho, \label{}\\
\left( \mu + \nu   \right)\left(\frac{\dot{a}^2}{a^2} + 2\frac{\ddot{a}}{a}   \right)     -\frac{1}{2}\left( \mu \dot{a}u_{\dot{a}} +\nu \dot{a}v_{\dot{a}} -\mu u -\nu v \right) - \frac{\mu}{6}a\left(\dot{u}_{\dot{a}}-u_a\right) - \frac{\nu}{6}a\left(\dot{v}_{\dot{a}}-v_a\right) &=& -p, \label{}\\
\dot{\rho} + 3H\left( \rho + p  \right) &=& 0,  \label{}
\end{eqnarray}
or equivalently:
\begin{eqnarray}
3\left( \mu + \nu \right) H^2 -\frac{1}{2}\left( \mu \dot{a}u_{\dot{a}} +\nu \dot{a}v_{\dot{a}} -\mu u -\nu v\right) &=& \rho, \label{AA1}\\
\left( \mu + \nu   \right)\left(2\dot{H} + 3H^2   \right)     -\frac{1}{2}\left( \mu \dot{a}u_{\dot{a}} +\nu \dot{a}v_{\dot{a}} -\mu u -\nu v \right) - \frac{\mu}{6}a\left(\dot{u}_{\dot{a}}-u_a\right) - \frac{\nu}{6}a\left(\dot{v}_{\dot{a}}-v_a\right) &=& -p, \label{AA2}\\
\dot{\rho} + 3H\left( \rho + p  \right) &=& 0.  \label{AA3}
\end{eqnarray}
The above system has two equations and five unknown functions, which are $a$, $\rho$, $p$, $u$ and $v$.\\
We now assume the following expressions for $u$ and $v$:
\begin{eqnarray}
u &=& \alpha a^n, \\
v &=& \beta a^m,
\end{eqnarray}
where $m$, $n$, $\alpha$ and $\beta$ are real constants. We also have that $u$ and $v$ can be expressed as:
\begin{eqnarray}
u &=& \alpha \left( \frac{v}{\beta}  \right)^{\frac{n}{m}}, \\
v &=& \beta \left( \frac{u}{\alpha}  \right)^{\frac{m}{n}}.
\end{eqnarray}
Then, the system made by Eqs. (\ref{AA1}), (\ref{AA2}) and (\ref{AA3}) leads to:
\begin{eqnarray}
3\left( \mu + \nu   \right)H^2 + \frac{1}{2}\left( \mu \alpha a^n + \nu \beta a^m  \right) &=& \rho, \label{23}\\
\left( \mu + \nu   \right) \left( 2\dot{H} +3H^2 \right) + \frac{\mu \alpha \left(n+3\right)}{6}a^n + \frac{\nu \beta \left(m+3\right)}{6}a^m &=& -p, \label{24}\\
\dot{\rho} + 3H\left( \rho + p  \right) &=& 0.    \label{24-1}
\end{eqnarray}
Finally, we have that the EoS parameter $\omega$ for this model is given by the relation:
\begin{eqnarray}
\omega =\frac{p}{\rho}=-1-\frac{2\left(\mu + \nu \right)\dot{H} -  \frac{\mu}{6}a\left(\dot{u}_{\dot{a}}-u_a\right) - \frac{\nu}{6}a\left(\dot{v}_{\dot{a}}-v_a\right)}  {3\left(\mu + \nu \right)H^2-\frac{1}{2}\left(\mu \dot{a}u_{\dot{a}} +\nu \dot{a}v_{\dot{a}}    -\mu u - \nu v  \right)}.
\end{eqnarray}

\section{INTERACTING DE IN $f\left(R, T\right)$ GRAVITY}
Solving the differential equation for $\rho_m$ given in Eq. (\ref{9}), we derive the following expression for $\rho_m$:
\begin{eqnarray}
\rho_m = \rho_{m0}a^{-3\left( 1+\delta  \right)}, \label{25}
\end{eqnarray}
where $\rho_{m0}$ indicates the present day value of $\rho_m$.\\
Using Eqs. (\ref{enerden}) and (\ref{25}) in the right hand side of the Eq. (\ref{23}), we obtain the following expression of $H^2$ as function of the scale factor:
\begin{eqnarray}
H^2&=& C_1a^{-\frac{\lambda + \sqrt{\lambda^2 - 8\varepsilon \left[\theta - 3\left(\mu + \nu \right)    \right]}}{2\varepsilon}}+ C_2a^{\frac{-\lambda + \sqrt{\lambda^2 - 8\varepsilon \left[\theta - 3\left(\mu + \nu \right)    \right]}}{2\varepsilon}} +\nonumber \\
&&\frac{\alpha \mu a^n}{n^2 \varepsilon + n\lambda  + 2\left[ \theta - 3\left(\mu + \nu   \right)   \right]} +
 \frac{\beta \nu a^m}{m^2 \varepsilon + m\lambda  + 2\left[ \theta - 3\left(\mu + \nu   \right) \right]} \nonumber \\
  &&-\frac{2a^{-3\left( 1+\delta  \right)}\rho_{m0}}{9\left( 1+\delta \right)^2\varepsilon    +2\theta - 3\left[ \lambda \left(1+\delta \right) +2\left( \mu + \nu  \right) \right]}, \label{26}
\end{eqnarray}
where $C_1$ and $C_2$ are two constants of integration.\\
In order to have a real and definite expression of $H^2$ given in Eq. (\ref{26}), the following conditions must be satisfied: $\varepsilon \neq 0$, $\lambda^2 - 8\varepsilon \left[ \theta - 3\left( \mu + \nu  \right)  \right] \geq 0$, $ n^2 \varepsilon + n\lambda  + 2\left[ \theta - 3\left(\mu + \nu   \right)   \right] \neq 0  $, $  m^2 \varepsilon + m\lambda  + 2\left[ \theta - 3\left(\mu + \nu   \right) \right]  \neq 0  $ and $9\left( 1+\delta \right)^2    +2\theta - 3\left[ \lambda \left(1+\delta \right) +2\left( \mu + \nu  \right) \right] \neq 0      $.\\
We can now derive the expressions of the first and the second time derivative of the Hubble parameter $H$, i.e. $\dot{H}$ and $\ddot{H}$, as functions of the scale factor $a$ differentiating  Eq. (\ref{26}) with respect to the cosmic time $t$:
\begin{eqnarray}
\dot{H}   &=&   - \frac{C_1}{2}  \left(\frac{\lambda + \sqrt{\lambda^2 - 8\varepsilon \left[\theta - 3\left(\mu + \nu \right)    \right]}}{2\varepsilon}  \right)  a^{-\frac{\lambda + \sqrt{\lambda^2 - 8\varepsilon \left[\theta - 3\left(\mu + \nu \right)    \right]}}{2\varepsilon}}+\nonumber \\
&&\frac{C_2}{2}\left(\frac{-\lambda + \sqrt{\lambda^2 - 8\varepsilon \left[\theta - 3\left(\mu + \nu \right)    \right]}}{2\varepsilon} \right) a^{\frac{-\lambda + \sqrt{\lambda^2 - 8\varepsilon \left[\theta - 3\left(\mu + \nu \right)    \right]}}{2\varepsilon}} +\nonumber \\
&&\frac{n\alpha \mu a^n}{2\left\{  n^2 \varepsilon + n\lambda  + 2\left[ \theta - 3\left(\mu + \nu   \right)   \right]\right\}} +
 \frac{m\beta \nu a^m}{2\left\{ m^2 \varepsilon + m\lambda  + 2\left[ \theta - 3\left(\mu + \nu   \right) \right]\right\}} \nonumber \\
  &&+\frac{3\left( 1+\delta  \right)a^{-3\left( 1+\delta  \right)}\rho_{m0}}{9\left( 1+\delta \right)^2\varepsilon    +2\theta - 3\left[ \lambda \left(1+\delta \right) +2\left( \mu + \nu  \right) \right]}    , \label{27}
\end{eqnarray}
\begin{eqnarray}
\ddot{H} &=&          H \left\{   \frac{C_1}{2}  \left(\frac{\lambda + \sqrt{\lambda^2 - 8\varepsilon \left[\theta - 3\left(\mu + \nu \right)    \right]}}{2\varepsilon}  \right)^2  a^{-\frac{\lambda + \sqrt{\lambda^2 - 8\varepsilon \left[\theta - 3\left(\mu + \nu \right)    \right]}}{2\varepsilon}}+ \right. \nonumber\\
&& \left. \frac{C_2}{2}\left(\frac{-\lambda + \sqrt{\lambda^2 - 8\varepsilon \left[\theta - 3\left(\mu + \nu \right)    \right]}}{2\varepsilon} \right)^2 a^{\frac{-\lambda + \sqrt{\lambda^2 - 8\varepsilon \left[\theta - 3\left(\mu + \nu \right)    \right]}}{2\varepsilon}} \right. +\nonumber \\
&&\left.\frac{n^2\alpha \mu a^n}{2\left\{  n^2 \varepsilon + n\lambda  + 2\left[ \theta - 3\left(\mu + \nu   \right)   \right]\right\}} +
 \frac{m^2\beta \nu a^m}{2\left\{ m^2 \varepsilon + m\lambda  + 2\left[ \theta - 3\left(\mu + \nu   \right) \right]\right\}} \right. \nonumber \\
  &&\left.-\frac{9\left( 1+\delta  \right)^2a^{-3\left( 1+\delta  \right)}\rho_{m0}}{9\left( 1+\delta \right)^2\varepsilon    +2\theta - 3\left[ \lambda \left(1+\delta \right) +2\left( \mu + \nu  \right) \right]}
  \right\}. \label{28}
\end{eqnarray}
Using Eqs. (\ref{26}), (\ref{27}) and (\ref{28}) in Eq. (\ref{enerden}), we obtain the following expression of the energy density $\rho_{DE}$:
\begin{eqnarray}
\rho_{DE}&=&   \frac{1}{2}\left[    \frac{\left(n^2 \varepsilon  +2 \theta + n\lambda  \right)\alpha \mu a^n}{\left\{  n^2 \varepsilon + n\lambda  + 2\left[ \theta - 3\left(\mu + \nu   \right)   \right]\right\}} +
 \frac{\left(m^2\varepsilon +2\theta + 2m\lambda   \right)\beta \nu a^m}{\left\{ m^2 \varepsilon + m\lambda  + 2\left[ \theta - 3\left(\mu + \nu   \right) \right]\right\}} \right. \nonumber \\
&&\left.             +6C_1\left(  \mu + \nu \right) a^{-\frac{\lambda + \sqrt{\lambda^2 - 8\varepsilon \left[\theta - 3\left(\mu + \nu \right)    \right]}}{2\varepsilon}} \right. \nonumber \\
&& \left.+ 6C_2\left(  \mu + \nu \right) a^{\frac{-\lambda + \sqrt{\lambda^2 - 8\varepsilon \left[\theta - 3\left(\mu + \nu \right)    \right]}}{2\varepsilon}}   \right.\nonumber \\
 &&    \left.  -\frac{2    \left[9\left( 1+\delta  \right)^2 \varepsilon +2\theta -3\left(  1+\delta  \right)\lambda\right]  a^{-3\left( 1+\delta  \right)}\rho_{m0}}{9\left( 1+\delta \right)^2 \varepsilon   +2\theta - 3\left[ \lambda \left(1+\delta \right) +2\left( \mu + \nu  \right) \right]}        \right].     \label{29}
\end{eqnarray}
Taking into account the expression of $\rho_{DE}$ given in Eq. (\ref{29}), we derive that the expression of the  pressure $p_{DE}$ of DE is given by:
\begin{eqnarray}
p_{DE}   &=&    C_1  \left(  \mu + \nu \right)   \left(\frac{ -6\varepsilon +    \lambda + \sqrt{\lambda^2 - 8\varepsilon \left[\theta - 3\left(\mu + \nu \right)    \right]}}{2\varepsilon}\right)  a^{-\frac{\lambda + \sqrt{\lambda^2 - 8\varepsilon \left[\theta - 3\left(\mu + \nu \right)    \right]}}{2\varepsilon}}\nonumber \\
&& C_2 \left(  \mu + \nu \right)   \left(\frac{ 6\varepsilon -    \lambda + \sqrt{\lambda^2 - 8\varepsilon \left[\theta - 3\left(\mu + \nu \right)    \right]}}{2\varepsilon}\right) a^{\frac{-\lambda + \sqrt{\lambda^2 - 8\varepsilon \left[\theta - 3\left(\mu + \nu \right)    \right]}}{2\varepsilon}}\nonumber \\
&& -\frac{\left[ 2\theta + n\left(n\varepsilon + \lambda   \right)  \right]\left(n+3  \right)\alpha \mu a^n}{6\left\{  n^2 \varepsilon + n\lambda  + 2\left[ \theta - 3\left(\mu + \nu   \right)   \right]\right\}} \nonumber\\
  &&-\frac{\left[ 2\theta + n\left(n\varepsilon + \lambda   \right)  \right]\left(m+3  \right)\beta \nu a^m}{6\left\{  m^2 \varepsilon + m\lambda  + 2\left[ \theta - 3\left(\mu + \nu   \right)   \right]\right\}} \nonumber\\
&&-\frac{    \left[9\left( 1+\delta  \right)^2 \varepsilon +2\theta -3\left(  1+\delta  \right)\lambda\right]  a^{-3\left( 1+\delta  \right)}\delta \rho_{m0}}{9\left( 1+\delta \right)^2\varepsilon    +2\theta - 3\left[ \lambda \left(1+\delta \right) +2\left( \mu + \nu  \right) \right]}            . \label{30}
\end{eqnarray}
Using the expressions of the energy density $\rho_{DE}$ and the pressure $p_{DE}$ of DE given, respectively, in Eqs. (\ref{29}) and (\ref{30}) and the expression of $\rho_m$ given in Eq. (\ref{25}), we get the EoS parameter $\omega_{DE}$ for DE and the total EoS parameter $\omega_{tot}$ as follow:
\begin{eqnarray}
\omega_{DE} &=& \frac{p_{DE}}{\rho_{DE}}, \label{31}\\
\omega_{tot} &=& \frac{p_{DE}}{\rho_{DE}+\rho_m}.\label{32}
\end{eqnarray}
We must remember here that we are considering the case of pressureless DM, so that $p_m=0$.\\
We now want to consider the properties of the deceleration parameter $q$ for the model we are considering. The deceleration parameter $q$ is generally defined as follow:
\begin{eqnarray}
q= -1 - \frac{a\ddot{a}}{\dot{a}^2} = -1 - \frac{\dot{H}}{H^2}, \label{33}
\end{eqnarray}
where the expressions of $H^2$ and $\dot{H}$ are given, respectively, in Eqs. (\ref{26}) and (\ref{27}).
The deceleration parameter, the Hubble parameter $H$ and the dimensionless energy density parameters $\Omega_{DE}$, $\Omega_m$ and $\Omega_k$ (which will be considered and studied in the following Sections) are a set of useful parameters if it is needed to describe cosmological observations.

\section{The statefinder parameters}
In order to have a better comprehension of the properties of the DE model taken into account, we can compare it with a model independent diagnostics which is able to differentiate between a wide variety of dynamical DE models, including the $\Lambda$CDM model. We consider here the diagnostic, also known as statefinder diagnostic, which introduces a pair of parameters $\left \{r, s\right \}$ defined, respectively, as follow:
\begin{eqnarray}
r &=& 1 + 3\frac{\dot{H}}{H^2}+ \frac{\ddot{H}}{H^3}=  1+\frac{3\dot{H}+\ddot{H}/H}{H^2}  , \label{34}\\
s&=& -\frac{3H\dot{H}+\ddot{H}}{3H\left( 2\dot{H}+3H^2  \right)}=  -\frac{3\dot{H}+\ddot{H}/H}{3\left( 2\dot{H}+3H^2  \right)}  . \label{35}
\end{eqnarray}
Using Eqs. (\ref{26}), (\ref{27}) and (\ref{29}), we get the statefinder parameters as:
\begin{eqnarray}
r &=& 1+ \frac{\rho_1}{\rho_2}, \label{36}\\
s &=& \frac{\zeta _1}{\zeta _2},\label{37}
\end{eqnarray}
with:
\begin{eqnarray}
\rho_1 &=&   C_1  \left(  \mu + \nu \right)\left(   \lambda + \sqrt{\lambda^2 - 8\varepsilon \left[\theta - 3\left(\mu + \nu \right)    \right]} \right) \times \nonumber\\
 &&  \left(\frac{ -6\varepsilon +    \lambda + \sqrt{\lambda^2 - 8\varepsilon \left[\theta - 3\left(\mu + \nu \right)    \right]}}{8\varepsilon^2}\right) a^{-\frac{\lambda + \sqrt{\lambda^2 - 8\varepsilon \left[\theta - 3\left(\mu + \nu \right)    \right]}}{2\varepsilon}}+ \nonumber \\
&& C_2 \left(  \mu + \nu \right)  \left(  - \lambda + \sqrt{\lambda^2 - 8\varepsilon \left[\theta - 3\left(\mu + \nu \right)    \right]} \right) \times \nonumber\\
 && \left(\frac{ 6\varepsilon -    \lambda + \sqrt{\lambda^2 - 8\varepsilon \left[\theta - 3\left(\mu + \nu \right)    \right]}}{8\varepsilon^2}\right) a^{\frac{-\lambda + \sqrt{\lambda^2 - 8\varepsilon \left[\theta - 3\left(\mu + \nu \right)    \right]}}{2\varepsilon}}+\nonumber \\
 &&\frac{\left(n+3  \right)n\alpha \mu a^n}{2\left\{  n^2 \varepsilon + n\lambda  + 2\left[ \theta - 3\left(\mu + \nu   \right)   \right]\right\}}+ \nonumber\\
  &&\frac{\left(m+3  \right)m\beta \nu a^m}{2\left\{  m^2 \varepsilon + m\lambda  + 2\left[ \theta - 3\left(\mu + \nu   \right)   \right]\right\}}- \nonumber\\
 &&\frac{   9  a^{-3\left( 1+\delta  \right)}\delta \left(1+\delta \right)\rho_{m0}}{9\left( 1+\delta \right)^2\varepsilon    +2\theta - 3\left[ \lambda \left(1+\delta \right) +2\left( \mu + \nu  \right) \right]},           \label{38} \\
\rho_2 &=& C_1a^{-\frac{\lambda + \sqrt{\lambda^2 - 8\varepsilon \left[\theta - 3\left(\mu + \nu \right)    \right]}}{2\varepsilon}}+ C_2a^{\frac{-\lambda + \sqrt{\lambda^2 - 8\varepsilon \left[\theta - 3\left(\mu + \nu \right)    \right]}}{2\varepsilon}} +\nonumber \\
&&\frac{\alpha \mu a^n}{n^2 \varepsilon + n\lambda  + 2\left[ \theta - 3\left(\mu + \nu   \right)   \right]} +
 \frac{\beta \nu a^m}{m^2 \varepsilon + m\lambda  + 2\left[ \theta - 3\left(\mu + \nu   \right) \right]} \nonumber \\
  &&-\frac{2a^{-3\left( 1+\delta  \right)}\rho_{m0}}{9\left( 1+\delta \right)^2\varepsilon    +2\theta - 3\left[ \lambda \left(1+\delta \right) +2\left( \mu + \nu  \right) \right]}, \label{39}
\end{eqnarray}
and:
\begin{eqnarray}
\zeta _1 &=& -   C_1  \left(  \mu + \nu \right)\left(   \lambda + \sqrt{\lambda^2 - 8\varepsilon \left[\theta - 3\left(\mu + \nu \right)    \right]} \right) \times \nonumber \\
&&   \left(\frac{ -6\varepsilon +    \lambda + \sqrt{\lambda^2 - 8\varepsilon \left[\theta - 3\left(\mu + \nu \right)    \right]}}{8\varepsilon^2}\right) a^{-\frac{\lambda + \sqrt{\lambda^2 - 8\varepsilon \left[\theta - 3\left(\mu + \nu \right)    \right]}}{2\varepsilon}}\nonumber \\
&& -C_2  \left(  \mu + \nu \right) \left(  - \lambda + \sqrt{\lambda^2 - 8\varepsilon \left[\theta - 3\left(\mu + \nu \right)    \right]} \right) \times \nonumber\\
&& \left(\frac{ 6\varepsilon -    \lambda + \sqrt{\lambda^2 - 8\varepsilon \left[\theta - 3\left(\mu + \nu \right)    \right]}}{8\varepsilon^2} \right)  a^{\frac{-\lambda + \sqrt{\lambda^2 - 8\varepsilon \left[\theta - 3\left(\mu + \nu \right)    \right]}}{2\varepsilon}}\nonumber \\
 &&-\frac{\left(n+3  \right)n\alpha \mu a^n}{2\left\{  n^2 \varepsilon + n\lambda  + 2\left[ \theta - 3\left(\mu + \nu   \right)   \right]\right\}} \nonumber\\
  &&-\frac{\left(m+3  \right)m\beta \nu a^m}{2\left\{  m^2 \varepsilon + m\lambda  + 2\left[ \theta - 3\left(\mu + \nu   \right)   \right]\right\}} \nonumber\\
 &&+\frac{   9  a^{-3\left( 1+\delta  \right)}\delta \left(1+\delta \right)\rho_{m0}}{9\left( 1+\delta \right)^2\varepsilon    +2\theta - 3\left[ \lambda \left(1+\delta \right) +2\left( \mu + \nu  \right) \right]},           \label{40} \\
\zeta_2 &=&   -       3C_1    \left(\frac{ -6\varepsilon +    \lambda + \sqrt{\lambda^2 - 8\varepsilon \left[\theta - 3\left(\mu + \nu \right)    \right]}}{2\varepsilon} \right) a^{-\frac{\lambda + \sqrt{\lambda^2 - 8\varepsilon \left[\theta - 3\left(\mu + \nu \right)    \right]}}{2\varepsilon}}\nonumber \\
&&+ 3C_2 \left(\frac{ 6\varepsilon -    \lambda + \sqrt{\lambda^2 - 8\varepsilon \left[\theta - 3\left(\mu + \nu \right)    \right]}}{2\varepsilon} \right) a^{\frac{-\lambda + \sqrt{\lambda^2 - 8\varepsilon \left[\theta - 3\left(\mu + \nu \right)    \right]}}{2\varepsilon}}\nonumber \\
 &&+3\frac{\left(n+3  \right)\alpha \mu a^n}{\left\{  n^2 \varepsilon + n\lambda  + 2\left[ \theta - 3\left(\mu + \nu   \right)   \right]\right\}} \nonumber\\
  &&+3\frac{\left(m+3  \right)\beta \nu a^m}{\left\{  m^2 \varepsilon + m\lambda  + 2\left[ \theta - 3\left(\mu + \nu   \right)   \right]\right\}} \nonumber\\
 &&+\frac{   18  a^{-3\left( 1+\delta  \right)}\delta \rho_{m0}}{9\left( 1+\delta \right)^2\varepsilon    +2\theta - 3\left[ \lambda \left(1+\delta \right) +2\left( \mu + \nu  \right) \right]}.                 \label{41}                 \label{41}
\end{eqnarray}

\section{Discussion}
In this Section, we discuss the behavior of the physical quantities
derived in the previous Sections. We have considered the following values for the parameters involved: $C_1 = 0.2$, $C_2 = 1.2$,  $m = 1.2,~n = 1.4,~\beta = 1.2,~\alpha = 1.5,~\theta = 0.002,~\lambda = 2,~\nu =
0.5,~\mu = 0.9,~\delta = 0.05~ \textrm{and}~\rho_{m0}=0.23$. We considered three different cases corresponding to three different values of the parameter $\varepsilon$, i.e. $\varepsilon=2,~3~\textrm{and}~4$.\\
In Figure 1, we plotted the expression of the Hubble parameter $H$, obtained from Eq. (\ref{26}), as function of the redshift $z$. It is evident that the Hubble parameter $H$ has a decaying behavior with varying values of the parameter $\varepsilon$ and the redshift $z$ going from higher to lower redshifts. However, this decaying pattern is apparent till $z\approx-0.5$. In fact, in a very late stage $z>-0.5$, it shows an increasing pattern.
\\
In Figure 2, we have plotted the time derivative of Hubble parameter
$\dot{H}$ against the redshift $z$. We have observed that for
$\varepsilon=3$, $\dot{H}$ transits from negative to positive side at
$z\approx-0.5$. However, for $\varepsilon=2~\textrm{and}~4$ this transition occurs at lower redshift $z\approx-0.1$.
In Figures 3 and 4 we have plotted, respectively, the equation of state (EoS) parameter for DE, defined as $\omega_{DE}=p_{DE}/\rho_{DE}$, and the effective EoS parameter, defined as $\omega_{eff}=p_{DE}/(\rho_{DE}+\rho_{DM})$, for the three different values of  $\varepsilon$ considered in this work. In Figure 3, we have observed that
for $\varepsilon=2$, $\omega_{DE}$ crosses the phantom divide $-1$ at
$z\approx 0$. For $\varepsilon=3$ the phantom divide is crossed at
$z\approx-0.2$. However, for $\varepsilon=4$, the equation of state
(EoS) parameter for DE stays below $-1$. Thus, for $\varepsilon=2$
and $3$, $\omega_{DE}$ transits from quintessence to phantom
i.e. has a quintom-like behavior. Instead, for $\varepsilon=4$, the EoS
parameter has a phantom-like behavior. In Figure 4, we have plotted the
effective EoS parameter $\omega_{eff}$. In this case, for all values of
$\varepsilon$ considered, there is a crossing of phantom divide. Moreover, for
$\varepsilon=4$, $\omega_{eff}$ crosses the phantom divide
 earlier respect to the other cases, in particular for  $z\approx0.2$. \\
The deceleration parameter $q$ has been
plotted as a function of $z$ in Figure 5. For $\varepsilon=2,~3$
there is a transition from positive to negative $q$, i.e.
transition from decelerated to accelerated expansion. For $\varepsilon=3$,
the deceleration parameter changes sign at $z=0$ and for
$\varepsilon=2$, it changes sign at $z\approx 0.1$. However, for
$\varepsilon=4$, the deceleration parameter always stays at negative level.
Thus, for $\varepsilon=4$, we are getting ever-accelerating universe.

Next, we have plotted in Figure 6
the fractional density of DE, given by $\Omega_{DE} =
\frac{\rho_{DE}}{3\widetilde{H}^2\left(z\right)}$, and the
fractional density of matter, given by $\Omega_{m} =
\frac{\rho_m}{3\widetilde{H}^2\left(z\right)}$  against the redshift $z$. $\widetilde{H}^2$ is defined as
$\widetilde{H}^2\left(z\right) = \left(\mu + \nu \right)H^2 +
\frac{1}{6}\left[ \alpha \mu \left(1+z \right)^{-n} + \beta \nu
\left(1 + z \right)^{-m} \right]$. The solid
lines correspond to $\Omega_{DE}$ and the dashed lines correspond
to $\Omega_{DM}$. In this Figure, there is a clear indication of
transition of the universe from dark matter dominated phase to the
dark energy dominated phase. At very early stage of the universe
$z>1$, the dark energy density is largely dominated by dark matter
density. We denote the cross-over point by $z_{cross}$ and it
comes out to be $z_{cross}\approx 0.5$, i.e. where
$\Omega_{DE}=\Omega_{DM}$ for all values of $\varepsilon$ considered in this work. Hence, the
$f(R,T)$ model, based on which we have reconstructed DE density, is
capable of achieving the present DE dominated universe
from the earlier dark-matter dominated universe.

Sahni et al. \cite{sah} recently demonstrated that the statefinder
diagnostic is effectively able to discriminate between different
models of DE. Chaplygin gas, braneworld, quintessence and cosmological constant models were investigated by Alam et al. \cite{alam}
using the statefinder diagnostics: they observed that the
statefinder pair could differentiate between these different
models. An investigation on statefinder parameters for
differentiating between DE and modified gravity was carried out in
\cite{wang}. Statefinder diagnostics for $f\left(T\right)$ gravity
has been well studied in Wu $\&$ Yu \cite{wu1}. In the $\left \{r,
s\right \}$ plane, $s > 0$ corresponds to a quintessence-like
model of DE and $s < 0$ corresponds to a phantom-like model of DE.
Moreover, an evolution from phantom to quintessence or inverse is
given by crossing of the fixed point $\left(r = 1, s = 0\right)$
in $\left \{r, s \right \}$ plane \cite{wu1}, which corresponds to
$\Lambda$CDM scenario. The statefinder parameter $\{r,s\}$ have
been plotted in Figure 7 for different values of the parameter $\varepsilon$. It is
clearly visible that the $\{r-s\}$ trajectories are
converging towards the fixed point $\{r=1,s=0\}|_{\Lambda
\textrm{CDM}}$. Thus, the $f(R,T)$ model is capable of attaining
the $\Lambda\textrm{CDM}$ phase of the universe. Furthermore, for
finite $r$, $s\rightarrow~-\infty$. Thus, the model can
interpolate between dust and $\Lambda$CDM phase of the universe.

\begin{figure}[h]
\includegraphics[width=16pc]{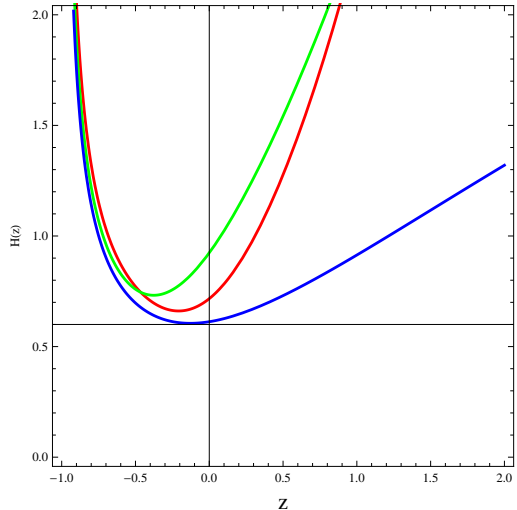}
\caption{\label{label}The Hubble
parameter $H$, obtained from Eq. (\ref{26}), as a function of redshift $z$. The red, green and blue
lines correspond to $\varepsilon=2,~3~\textrm{and}~4$, respectively.}
\end{figure}

\begin{figure}[h]
\includegraphics[width=16pc]{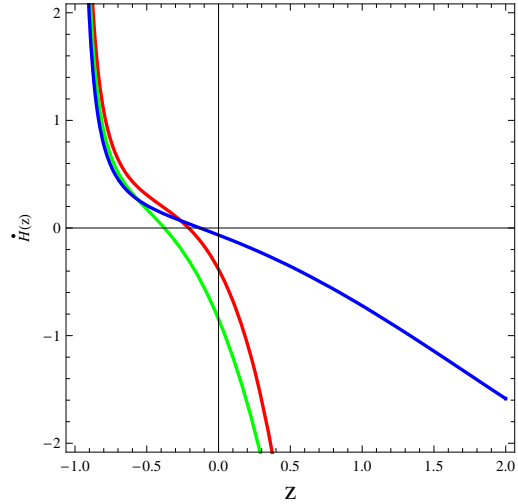}
\caption{\label{label}The time derivative of reconstructed Hubble
parameter as a function of redshift $z$. The red, green and blue
lines correspond to $\varepsilon=2,~3~\textrm{and}~4$ respectively.}
\end{figure}

\begin{figure}[h]
\includegraphics[width=16pc]{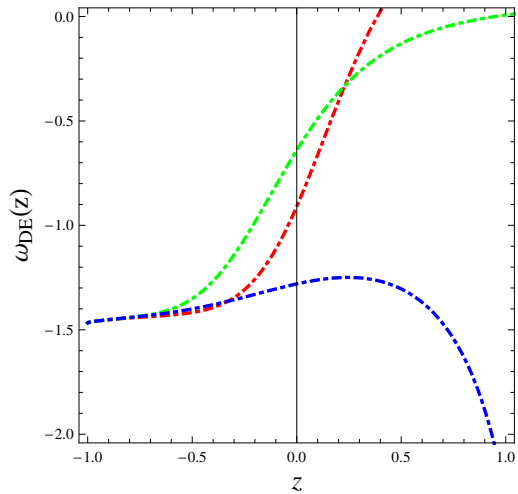}
\caption{\label{label}The EoS parameter $\omega_{DE}$ for the
reconstructed DE.}
\end{figure}

\begin{figure}[h]
\includegraphics[width=16pc]{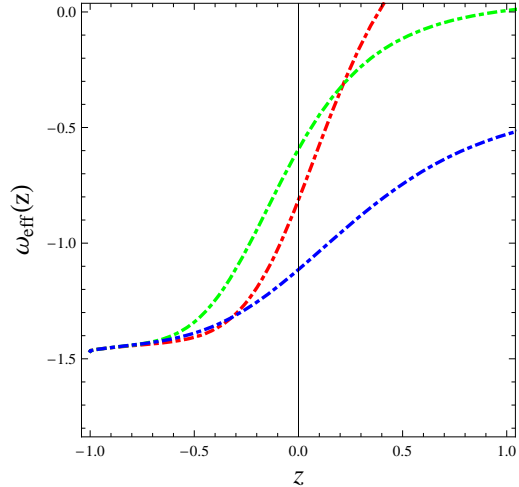}
\caption{\label{label}The effective EoS parameter
$\omega_{eff}(z)=\frac{p_{DE}}{\rho_{DE}+\rho_{DM}}$.}
\end{figure}

\begin{figure}[h]
\includegraphics[width=16pc]{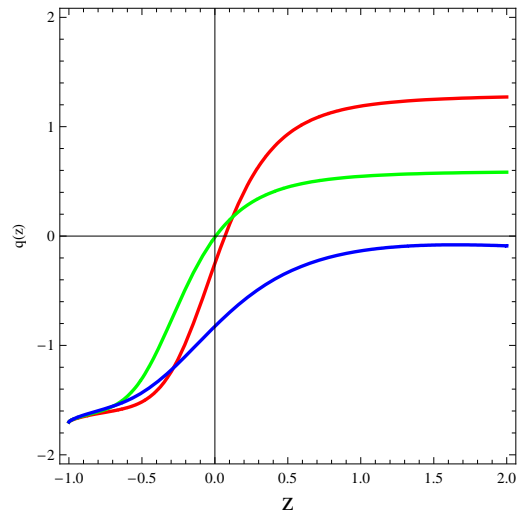}
\caption{\label{label}The deceleration parameter $q$ as a function
of $z$.}
\end{figure}

\begin{figure}[h]
\includegraphics[width=16pc]{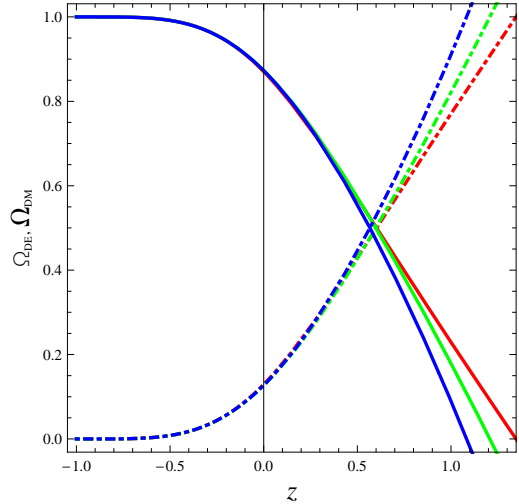}
\caption{\label{label}The fractional densities $\Omega_{DE}$
(smooth lines) and $\Omega_{DM}$ (dashed lines) as function of
redshift $z$.}
\end{figure}

\begin{figure}[h]
\includegraphics[width=16pc]{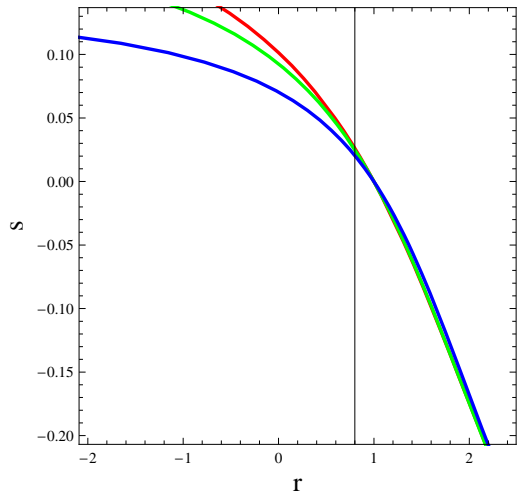}
\caption{\label{label} Statefinder trajectories for various choices
of parameters.}
\end{figure}

\begin{figure}[h]
\includegraphics[width=16pc]{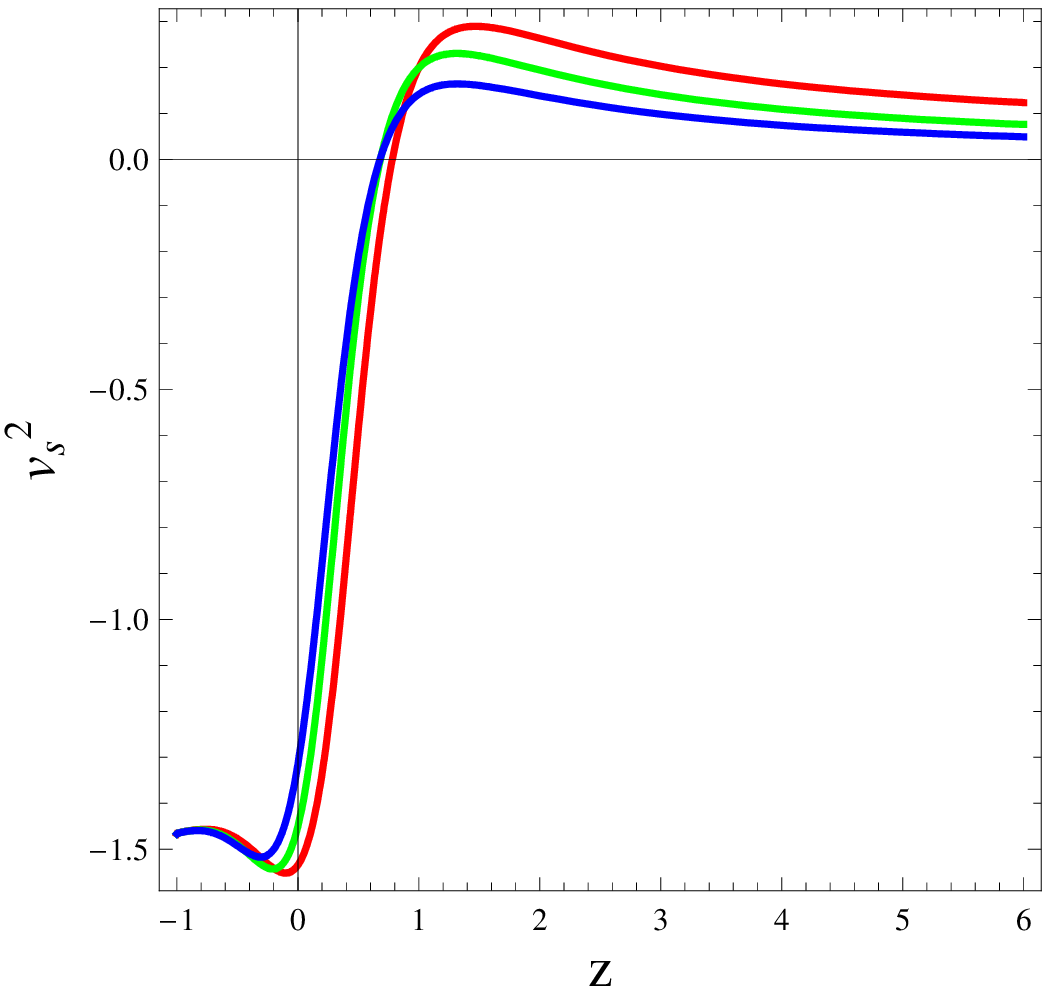}
\caption{\label{label}Squared speed of sound
$v_s^2=\frac{\dot{p}}{\dot{\rho}}$ as a function of redshift $z$.}
\end{figure}

Finally, in Figure 8, we plotted the squared speed of the sound,
defined as $v_{s}^{2}=\frac{\dot{p}}{\dot{\rho}}$, where the upper
dot indicates derivative with respect to the cosmic time $t$ for the model we are considering as a function of $z$. The sign of the squared speed of sound is fundamental in order to study the stability of a background evolution. A negative value of $v_s^2$ implies a classical instability of a given perturbation in general relativity \cite{vs1,vs2}. Myung \cite{vs2} recently observed that the squared speed of sound for HDE stays always negative if the future event horizon is considered as IR cutoff, while for Chaplygin gas and tachyon $v_s^2$ is observed non-negative. Kim et al. \cite{vs1} found that $v_s^2$ for
Agegraphic DE (ADE) stays always negative, which leads to the instability of the perfect fluid for the model. Moreover,
it was found that the ghost QCD \cite{vs3} DE model is unstable. In a recent work, Sharif $\&$ Jawad \cite{vs4} have
shown that interacting new HDE is characterized by negative squared speed of the sound. \\
In the recent work of Pasqua et al. \cite{miofrt}, authors observed that the interacting Modified Holographic Ricci DE (MHRDE) model in $f\left(R,T\right) = \mu R + \nu T$ gravity is classically stable. \\
Jawad et al. \cite{miovs1} was shown that $f\left(G\right)$ model in HDE scenario with power-law scale factor is classically unstable.\\
Pasqua et al. \cite{miovs2} showed  that the DE model  based on Generalized Uncertainty Principle (GUP) with power-law form of the scale factor $a\left(t\right)$ is instable.\\
We have observed that, for all values of $\varepsilon$, $v_s^2$ is positive
till the redshift $z\approx0.5$. However, after this stage it enters the
negative region. Thus, although the model is classically stable in
the early universe, for present universe it is classically
unstable.

\section{Concluding remarks}
In this work, we have considered a recently proposed model of
energy density of DE which depends by three terms, one
proportional to the squared Hubble parameter $H$, one proportional
to the time derivative of $H$ and one proportional to the second
time derivative of $H$ interacting with pressureless DM in the
framework of the $f\left(R,T\right)$ modified gravity theory for
the special model given by $f\left(R, T\right) = \mu R + \nu T$,
where $\mu$ and $\nu$ represents two constants. The DE model
considered here reduces to other two well-studied DE model (the
Ricci DE model and the DE energy density model with
Granda-Oliveros cut-off) for some particular values of the three
parameters involved, i.e. $\varepsilon$, $\lambda$ and $\theta$.
We have derived the expressions and studied the behavior of some
important physical quantities which gave useful hints about the
model studied. The Hubble parameter $H$ exhibits a decaying behavior
going from higher to lower redshifts until the redshift of about $z\approx-0.5$, when it starts to increase. The time derivative of the Hubble parameter, i.e. $\dot{H}$,
shows a transition from negative to positive values for different values of the redshift $z$ according to
the value of $\varepsilon$ considered.
We have observed that the equation of state (EoS)
parameter of DE exhibits a transition from $\omega<-1$ to $\omega>-1$
i.e. transition from quintessence to phantom (i.e. quintom) with
the evolution of the universe for $\varepsilon =2$ and $\varepsilon =3$, instead it is always negative for
$\varepsilon =4$. Moreover, the effective equation of state (EoS) parameter $\omega_{eff}$ always shows a transition from quintessence to phantom. Hence, we can conclude that the reconstructed $DE$ model
based on the $f\left(R,T\right)$ gravity model considered leads to a equation of
state (EoS) parameter that has a quintom-like behavior. We have further
observed that the said model is capable of attaining the dark
energy dominated accelerated phase of the universe from dark model
dominated decelerated phase of the universe. Through statefinder
trajectories we have shown that the $f(R,T)$ is capable of
attaining the $\Lambda$CDM phase of the universe and can
interpolate between dust and $\Lambda$CDM phase of the universe.
Through squared speed of sound we have seen that the model under
consideration is classically stable in the earlier stage of the
universe, but classically unstable in the current stage.

\section{Acknowledgement}
The second author acknowledges financial support from the
Department of Science and Technology, Govt. of India under project
grant no. SR/FTP/PS-167/2011.

\end{document}